\documentclass[conference]{IEEEtran}
\IEEEoverridecommandlockouts
\usepackage{cite}
\usepackage{amsmath,amssymb,amsfonts}
\usepackage{algorithmic}
\usepackage{graphicx}
\usepackage{textcomp}
\usepackage{xcolor}
\usepackage{subcaption}
\usepackage{array}
\usepackage{graphicx}
\usepackage[export]{adjustbox}
\usepackage{tikz}
\usepackage{float}
\usetikzlibrary{arrows}
\usepackage{verbatim}
\usepackage{comment}
\usepackage{bibentry}
\usepackage{graphicx}
\usepackage{microtype}
\graphicspath{Chapters/Figures}
\def\BibTeX{{\rm B\kern-.05em{\sc i\kern-.025em b}\kern-.08em
		T\kern-.1667em\lower.7ex\hbox{E}\kern-.125emX}}
\begin{document}
	\title{An Intelligent and Time-Efficient DDoS Identification Framework for Real-Time Enterprise Networks \\
    \large SAD-F: Spark Based Anomaly Detection Framework
    }
	\author{\IEEEauthorblockN{Awais Ahmed}
		\IEEEauthorblockA{
			NUCES-FAST\\
			k163053@nu.edu.pk}
		\and
		\IEEEauthorblockN{Sufian Hameed}
		\IEEEauthorblockA{
			NUCES-FAST\\
			sufian.hameed@nu.edu.pk}
		\and
		\IEEEauthorblockN{Muhammad Rafi}
		\IEEEauthorblockA{
			NUCES-FAST \\
			muhammad.rafi@nu.edu.pk}
		\and
		\IEEEauthorblockN{Qublai Khan Ali Mirza}
		\IEEEauthorblockA{
			University of Gloucestershire\\
			qalimirza@glos.ac.uk}
	}
\maketitle
\begin{abstract}

Enterprise networks face a magnitude of threats that are managed and mitigated with a combination of proprietary and third-party security tools and services. However, the techniques and principles employed by the said tools, techniques and services are quite conventional and lack the rapid evolution, as required to protect against modern, state-of-the-art threats faced, specifically, against distributed denial of service (DDoS) attacks. The lack of efficiency of a network is directly proportional to the number of applications and services it hosts, particularly to protect against external and internal threats. Moreover, the effectiveness of such security mechanisms relies on their independent and proactive approach, which is only as effective as their knowledge of known malware and their attack vectors that becomes obsolete when there is a new malware or a zero-day vulnerability is exploited. 

This paper presents an intelligent, highly responsive, and scalable security framework for enterprise networks. The proposed framework incorporates Apache Spark Framework for security analytics and accurately identifies anomalies specifically pertaining to DDoS attacks from real-time network traffic by using customised machine learning algorithms meticulously trained against selected feature-set. The results are tested against different scenarios and bench-marked with the results achieved by related studies in similar scenarios.

\end{abstract}
\begin{IEEEkeywords}
DDoS, Network Security, Apache Spark, Anomaly Detection, Machine Learning, Big Data Analytics, Security Analytics, Malware.
\end{IEEEkeywords}
\section{Introduction}

The heterogeneity in modern enterprise networks has enabled organizations to not only expand their business operations, it has also provided an opportunity for businesses to efficiently use their skilled resources in order to optimise the effectiveness of every operation. The inclusion of diverse set of devices and tools in regular enterprise networks, has significantly increased the efficacy of majority of related operations. This level of heterogeneity has also diminished physical and geographical boundaries generally faced by enterprises in the past \cite{schleich_performance_2017}. 

The benefits of loosely coupled infrastructure for modern enterprises outweighs the regular network policies and compliance, followed by such enterprises to enhance performance and security \cite{li_methods_2018}. However, securing such a loosely coupled enterprise network is significantly complicated and requires a combination of tools and techniques that can detect and prevent proactively with minimum to no human dependency. 

In current security ecosystem, if an organization is infected, it takes approximately six months to just identify that infection \cite{osborne_most_2015}, and that too, if there is a predefined malicious behaviour identified by intrusion detection systems (IDS). Generally, the technique of exploiting enterprise networks without triggering security alerts with malicious activities, is used by attackers with an intent of distributed denial of service (DDoS) attacks \cite{tariq_collaborative_2011}.

To timely detect DDoS or similar attacks, enterprises heavily rely on tools that are based on security analytics techniques, which means identification based on; signatures, heuristics, and behaviours of known malware, along with their attack vectors \cite{ashraf_handling_2014}. This approach is quite effective and can protect against numerous attacks, as the majority of malware released every day are variants of previously known malware \cite{moustafa2015unsw}. Therefore, using signature, heuristics, and brief behavioural feature-set, is quite effective. However, if the attack is originated by a new/unknown malware, exploiting a zero-day vulnerability in a network, then it can bypass the said security measures without getting detected. 

The effectiveness of Intrusion detection systems, antiviruses, and other similar security tools, against unknown malware, has been practically evaluated a number of times and the results are quite different from what is claimed by vendors of said tools. The lack of threat and anomaly detection is not the only issue, the network and computing resources consumed by the network security tools, along with the level of privilege they require, gives them the highest level of priority on the infrastructure. According to a recent evaluation, enterprise antiviruses consume 45\% of CPU resources, while scanning for threats and they are able to detect less than 60\% of threats \cite{mirza_cloud-based_2016}. These statistics quite accurately depict the performance of security tools, along with the implications they can have on host networks.

\begin{figure}
    \centering
    \includegraphics[width=0.9\linewidth]{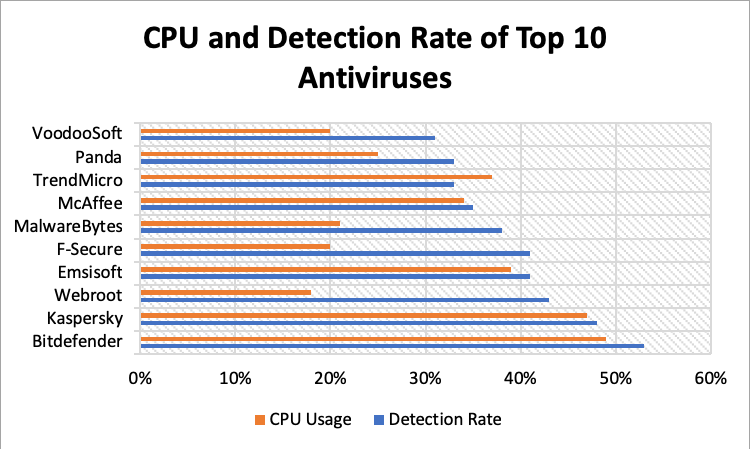}
    \caption{Antivirus Evaluation w.r.t. CPU Usage and Detection Rate \cite{mirza_cloud-based_2016}}
    \label{fig:my_label}
\end{figure}

The issues related to network security tools are not recently identified, it’s an on-going discussion in the security community. There are several solutions proposed by different researchers that quite accurately detect previously unknown malware using novel techniques. Additionally, the implementation of machine learning (ML) algorithms, artificial neural networks, etc.  to accurately detect malicious activity is quite successfully proposed and evaluated by several studies. In some selected studies, the level of accuracy for malware identification is 97\% - 98\% \cite{ali_mirza_cloudintell_2018}, which is significantly higher than the conventional security tools.

Moreover, similar machine learning techniques have been used by other selected studies, discussed in later sections, to accurately detect malicious activities in network logs, pertaining to DDoS and other similar attacks. The said approaches train ML algorithms against heuristics to accurately identify anomalies in network logs. The results of these techniques also surpass the identification statistics associated with conventional security tools. 

The aforementioned techniques proposed and self-evaluated by many studies, depict promising results while accurately detecting malware in a single machine or anomalies in network logs, associated with DDoS. However, there are two critical aspects of enterprise network security that are quite commonly overlooked; live anomaly detection and energy efficiency. 

The aforementioned proposed solutions do provide techniques with promising detection rate and level of accuracy but rarely evaluate the solution on the basis of energy efficiency. The effectiveness of such solutions is directly proportional to the resources they consume, however, in real-time network environments it is required to be inversely proportional. Moreover, the higher detection rate is achieved against logged data, which is quite different environment as compared to live traffic. A majority of the studies that claim the higher level of accuracy, deal with the stored feature-set, network logs, etc. Moreover, the primary difference in a live network and in stored network logs is the level of entropy the live environment entails. This means, if machine learning algorithms are trained on stored network logs, then it is quite likely that their implementation on the live network will generate a high number of false positives and negatives.

There exists a significant gap in security mechanisms that are currently used and even proposed to fill the existing gaps in network security. It is required to have a solution that independently and accurately identifies threats and anomalies in live network traffic with minimum amount of time. 

This paper initially presents a comparative analysis for streaming NetFlow traffic, including KDD, and UNSW-NB-15, which is later used for training supervised machine learning models for early anomaly detection. Moreover, the UNSW-NB-15 NetFlow dataset is then incorporated with Hadoop Framework: Spark, the results of this integration are compared with existing Hadoop Framework: MapReduce. This comparative analysis presents a significant improvement of real-time data analysis w.r.t. time and accuracy, by accurately detecting anomalies in a very short interval.

The framework, SAD-F: Spark-based Automated Anomaly Detection Framework, proposed in this paper, is specifically designed to befit real-time network security scenarios, namely; anomaly detection in live enterprise networks within a very short interval and anomaly identification through analysis of stored indicators of compromise (IoC). These two primary features of the proposed framework are the key elements that address the aforementioned gap in the research in the area of network security. Moreover, to ensure that the proposed framework meets the performance parameters initially set, a thorough evaluation is performed. Additionally, the results of evaluation are used to compare with the results of existing solutions and benchmarking. 

SAD-F is more appealing and close to real-time network security requirements, as compared to similar approaches for the following reasons: 
\begin{itemize}
\item   Although the frameworks, solutions, tools, and techniques currently used in conventional enterprise network environments, and proposed to fill the gaps in the existing security solutions, do address the issues. However, the current solutions lack the incorporation of entropy associated with the real-time network traffic, and the approach incorporating real-time entropy to achieve high level of accuracy has a lasting and scalable impact.
\item   Enterprise networks have a requirement of immediate identification of anomalies that can lead to malicious entities in the network, specifically, the ones leading to DDoS. The incorporation of security analytics, supported by machine learning models, trained against customised feature-set, enable SAD-F to identify anomalies in live network traffic stream in a responsive manner. Following are the machine learning algorithms that support the proposed framework.
\begin{itemize}
\item   K-Nearest Neighbour (KNN)
\item   Naïve Bayes (NB)
\item   Random Forest (RF)
\item   Decision Tree (DT)
\item   Support Vector Machine (SVM)
\end{itemize}

\item  The response time offered by SAD-F is benchmarked against existing solutions, which means that the effectiveness of the proposed framework is measured against accuracy and timeliness. These are the two parameters that complement each other while identifying the effectiveness of such a framework. 
\item  The proposed framework is implemented using open source highly responsive analytics engine designed for large-scale data processing, known as Spark. Therefore, the results presented, discussed, and benchmarked later in this paper are not limited to the experimental dataset, it is also highly scalable to accommodate any form of packet stream in a real-time heterogeneous network. 
\end{itemize}

The rest of the paper is structured as follows; Section \ref{RL} presents a discussion on the research performed in the same area, Section \ref{SAD-F} presents an in-depth discussion on the proposed framework and how different aspects of the framework are implemented, Section \ref{datasets} presents a discussion on the datasets used in the experiments and benchmarking, Section \ref{datasets} presents a step-by-step walkthrough of testbed design, Section \ref{experiments} is comprised of experiments performed, analysis of results, along with the critical evaluation of the proposed framework and benchmarking based on related research results, Section \ref{conclusion} presents the conclusion of the study.

\section{Related Work} \label{RL}
There are some notable solutions published on network anomaly detection techniques \cite{karimi2016distributed},including studies incorporating machine learning in their solutions. There are also such studies that focus on anomaly detection in big-data related domains \cite{casas2017network,velea2017feature}. In recent studies, many different techniques have been studied to solve the traffic classification problem. The majority of current classification approaches still rely on packet header based anomaly detection and protocol based anomaly detection \cite{davis2011data}.

In paper \cite{fontugne2010mawilab} Romain et al. proposed a new framework; Hashdoop (extension to Hadoop MapReduced). Hashdoop suggest a solution over a drawback of Hadoop (Spatial and Temporal structure) when it comes to the point of splits of network traffic over different nodes for distributed processing. Hashdoop splits traffic with hash function to preserve traffic structure which lead to outstanding performance over detection of network anomalies. Hashdoop were evaluated with two anomaly detectors and fifteen traces of Internet backbone traffic captured between 2001 and 2013. Hashdoop with 6-node cluster increased the throughput and enable real-time detection of large analyzed traces. Hashdoop improves the overall accuracy for anomalies detection process. 

Hashdoop, experiments were conducted on MAWI archive dataset and more precisely, traffic captured at between U.S and Japan on Sample point B and Sample point F. There is a performance trade-off, authors noted that dividing traffic into many small splits can cause for adverse results because each split may contain insufficient traffic for the statistical analysis, they left this gap for future work or considered to be an open challenge. 
Taking the above stated gap as our research study, we are expecting a step ahead solution; instead of using MapReduced technique we will utilize distributed computing stream processing for analysis of the same data and bears better results from existing study.

In paper \cite{dromard2015unsupervised} Juliette et al. stated in their study, the problem of unsupervised network anomaly has been studied in the last decade. Many studies were proposed time to time. Furthermore, they commented mostly unsupervised network anomaly detectors were depend on clustering techniques. Clustering algorithms group similar flows on to same cluster. The study presented a novel technique, unsupervised Network Anomaly Detectors Analysis (UNADA). This took advantage of distributed computing system to speed up their process. Dividing the features space in sub spaces which allows UNADA to run in parallel multiple DBSCANs and EA algorithms. Lastly, authors claimed that their proposed techniques' UNADA's deployment in real time stream tool 'Spark' can improve execution time by a factor of 13. Experiments were conducted on grid’5000 testbed.

Extensive work have been proposed in the field of SDN for DDoS mitigation. In paper \cite{hameed2017leveraging} authors proposed a collaborative distributed denial of service (DDoS) attack mitigation scheme using software defined network. They design a secure controller to controller (C-to-C) protocol that allows software defined network controllers lying in different autonomous systems (AS) to securely communicate and transfer attack information with each other. Further, as in extension they proposed an other scientific work \cite{hameed2018sdn}. They introduced three different deployment approaches i.e., linear, central and mesh. Which enables efficient notification along the path of an ongoing attack and effective filtering of traffic near the source of attack, thus saving valuable time and network resources. In experiments, they showed SDN based collaborative scheme is capable of efficiently mitigating DDoS.

In paper \cite{karimi2016distributed} Ahmad et al.presented in their work, one of the challenging module of the intrusion detection system known as feature extraction, and presented comparative analysis/results for TCP based traffic. They implemented their system by using Apache Spark and Netmap. Authors claimed, that their proposed system works well for small organization. This system was designed on the top of CAIDA NetFlow attack dataset. Limitation of their work can be overlook by deploying same system for large scale NetFlow traffic dataset, specifically for large organizations.  

In paper \cite{vcermak2016performance} Milan et al. proposed a novel performance benchmark based on common security analysis algorithms for NetFlow data to show suitability of distributed stream processing system. Hadoop based system works only for batch processing and do not offer stream processing (real-time) analysis.

In order to overcome the limitations of Hadoop, three of the most used distributed systems were taken for stream processing experiments such as Spark, Samza, and Storm. Experiments were conducted on the top of CAIDA NetFlow dataset. Lastly, this study presents performance benchmark chart to justify the outcome.

In paper \cite{casas2017network} and \cite{casas2016big} Pedro Crass et al. introduced a new framework for network analytic named BIG-DAMA, a Big Data Analytic framework (BDAF) for network traffic monitoring and analysis applications (NTMA). BIG-DAMA is a new and flexible framework, which is capable to analyze and store large data including stream and batch. This study also implements multiple data analytic techniques for network security and anomaly detection. Different machine learning were used. Authors claimed that they applied their technique with different types of attacks and benchmark outcome. Experiments were conducted on the top of MAWI dataset. Further, authors claimed that BIG-DAMA can speed up computation by a factor of 10 with respect to existing solutions. Proposed system; BIG-DAMA can easily deployable in cloud environment using virtualization techniques.

After going through from related work, we came up with a system and expects efficient in terms of time of computation and accurate enough from existing system. In order to achieve the objective, we deployed two of the streaming dataset including KDD99 and UNSW-NB15 and predicting accurate results for anomalies such as DoS, DDoS, IP Scanning, Port Scanning, etc. This paper investigates existing and proposed work and benchmark the difference. Lastly, this paper also list literature work on the Netflow data preprocessing techniques.

\section{Spark based Anomaly Detection Framework - (SAD-F)} \label{SAD-F}
The purpose of this study is to contribute and to fill the research gap as highlighted from \cite{hameed2016efficacy,hameed2018hadec,casas2017network} discussed above in section Problem Statement. The follow-up sections describes the subsections including proposed framework, dataset selection, data preprocessing techniques, model selection, results, conclusion and future work.
\begin{figure*}[t]
    \includegraphics[width=1.0\linewidth]{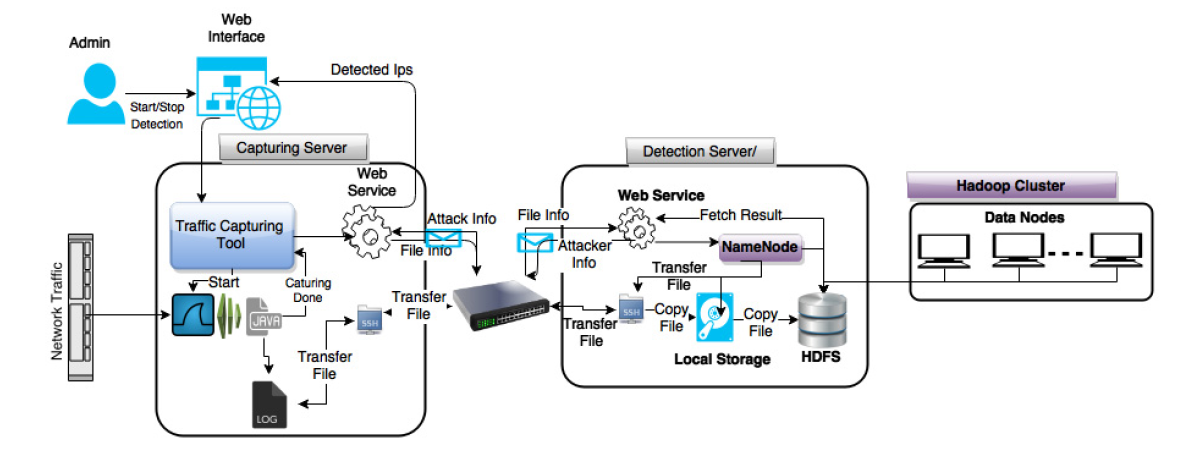}
    \caption {HADEC: Hadoop based Live DDoS Detection framework;{"figure reused from scientific paper" \cite{hameed2016efficacy,hameed2018hadec}}}
    \label{fig:HADEC}
\end{figure*}
\begin{figure*}[t]
    \centering
    \includegraphics[width=1.0\linewidth]{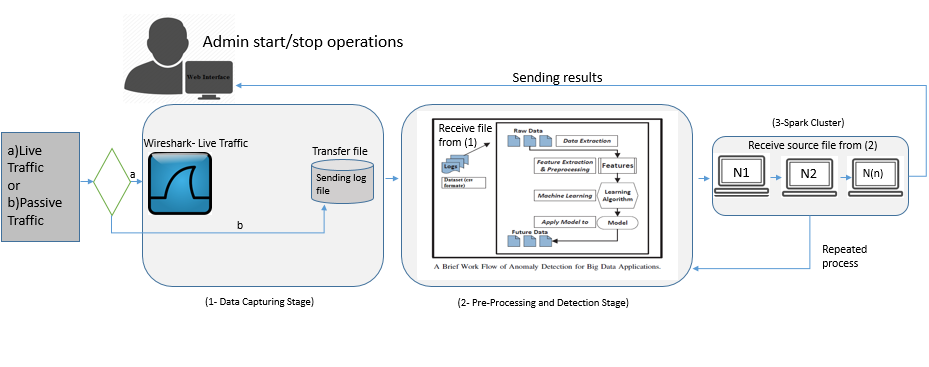}
    \caption  {SAD-F: Spark Based Anomaly Detection Framework}
    \label{fig:SAD-F}
\end{figure*}
\subsection{Proposed Framework}
In this study, we propose a successor of our previously purposed framework in base study \cite{hameed2016efficacy,hameed2018hadec}, which comprise of five major phases implemented as separate components including:
\begin{enumerate}{
    \item Netflow traffic capturing server and or netflow traffic selector
    \item Preprocessing of netflow data
    \item Spark cluster simulation phase
    \item Anomaly detector
    \item Result in a form of graph or notification to the concern body}
\end{enumerate}
As per limitations discussed, we upgraded framework \ref{fig:HADEC} adding stream processing by spark and removing counter based algorithm which were supposed to find listed anomalies. But this proposed SAD-Framework is mainly focused towards finding anomalies from known as well as unknown data with preferred low-level hardware also known as commodity hardware. One can see difference between proposed framework components and previous work \ref{fig:HADEC} and \ref{fig:SAD-F}.

\subsection{Netflow traffic capturing server and or netflow traffic selector}
In this phase, one may supposed to directly pass Netflow dataset or captures live traffic with the help of network capturing tool Wireshark with predefined filters to filter network traffic.

This framework is flexible and offers admin perspective web interface through which admin can tune capturing process by parameter tuning for desired live traffic. But at this stage we are not designing admin portal instead we will use passive netflow data to the file transfer stage. After loading capture file, file format could be (.cap, .pcap or csv) this file will pass over "Preprocessing and Detection stage" which is further discussed in upcoming subsection.

\begin{figure*}[t]
    \centering
    \includegraphics[width=0.9\linewidth]{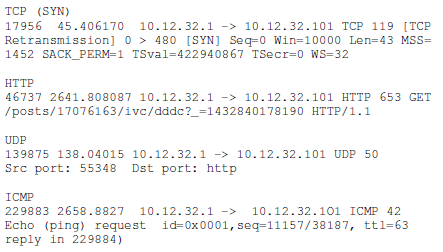}
    \caption{Wireshark (T-Shark)- Example filters for Live Capturing;{"figure reused from scientific paper" \cite{hameed2016efficacy,hameed2018hadec}}}
    \label{fig:filters}
\end{figure*}

Live traffic may contain some of predefined filters to capture required network traffic. For example, parameters could be packet size, no of frames and desired protocol filter etc. As admin finalized the parameter-tuning/ filter-setting capturing tool begins to capture as depicted in figure-\ref{fig:SAD-F}. Along with the traffic capturing process that file will be logged and will be transferred further for preprocessing and detection process which is further discussed below in subsection. Example filters are shown in figure-\ref{fig:filters} as reference these filters may help future researchers to begin with live capturing.

\subsection{Data preprocessing techniques}
The hard part of the anomaly detection process is the preprocessing of network traces. Data preprocessing is recognized as an important step in anomaly detection. Review shows that preprocessing needs domain experts for better features selection which leads improvements in results. The review also finds that many studies limit their preprocessing steps and do not discuss in details in their work. \cite{casas2017network}

As network traffic file will arrive at this stage preprocessing and anomaly detection process will begin. Data preprocessing is a must step in all knowledge discovery tasks from which preprocessing of network-based intrusion detection took 50 percent \cite{davis2011data} effort of the overall process which results better to classify network traffic as normal or anomalous. Removing strong co-relate,irrelevant and redundant features also improves the detection rate for learning based algorithms. This study focuses on the netflow preprocessing stage and for that we designed a novel approach for preprocessing and we studied various formal process models which were proposed for knowledge discovery and data mining (KDDM) as discussed in \cite{kurgan2006survey}. After thoroughly reviewing studies \cite{davis2011data, kurgan2006survey}, we found standard preprocessing steps include \textbf{dataset creation, data cleaning, integration, feature construction to derive new higher-level features, feature selection to choose the optimal subset of relevant features, reduction.} The most relevant steps for network intrusion deduction systems are briefly described:
\begin{itemize}
    \item \textbf{Dataset creation:}  Involves identifying representative network traffic for training and testing. These datasets should be labeled indicating whether the connection is normal or anomalous. Labeling network traffic can be a very time consuming and difficult task.
    \item \textbf{Feature construction:} It aims to create needed additional features with a better discriminating ability than the initial feature set. This step can bring significant improvement to machine learning algorithms. Features can be constructed manually, or by using data mining methods such as sequence analysis, association mining, and frequent episode mining
    \item \textbf{Reduction:}  It is commonly used technique to decrease the dimensionality of the dataset by discarding any redundant or irrelevant features. This process of feature optimization is called feature
    selection, and is commonly used to alleviate “the curse of dimensionality”. Data reduction can also be achieved with feature extraction which transforms the initial feature set into a reduced number of new features. Principal component analysis (PCA) is a common linear method used for data reduction.
\end{itemize}
Preprocessing converts network traffic into a series of observations, where each observation is represented as a feature vector. Observations are optionally labeled with it's classes, such as “normal” or “anomalous”. These feature vectors are then suitable as input to data mining or machine learning algorithms. Machine learning is the use of algorithms which evolve according to the labeled data instances (observations) provided to it. The algorithms are able to generalize from these observations, hence allowing future observations to be automatically classified. Machine learning is widely used in anomaly-based NIDS with examples including \cite{mahoney2001phad} and the Principal Component Classifier by \cite{shyu2003novel}

After successful process, the preprocessed file will be transferred further to spark cluster manager for further classification process and file will be tested against pre-trained model and results will be logged for further notification process which is further discussed in detail in upcoming section.
\subsection{Anomaly Detection Techniques: Network Intrusion Detection
Systems}
To the best of literature review, we are concluding that intrusion prevention techniques are imperfect, monitoring for security compromises is required. This is the basic role of network intrusion detection systems (IDSs). Our proposed NIDS aim to detect malicious activity in near *real-time and raise an alert.

Anomaly detection in intrusion based detection systems provides a way to detect a number of attacks, which results network failures at real time or be the basis for future mishaps. \textbf{Curing anomaly detection at early stages are preferable due to many factors including cost, response time of system, customer satisfaction.} 

In this paper, we are utilizing following listed supervised techniques for anomaly detection.
\begin{itemize}
    \item k-nearest neighbors algorithm
    \item Naive bayes
    \item Random forest 
    \item Decision Tree 
    \item Support Vector Machine
\end{itemize}
In this section we are not supposed to discuss models details in perspective of machine learning although interested readers are redirected to these \cite{Top10,algo}.
\subsection{Spark cluster simulation phase}
According to our architecture, after preprocessing of dataset, Spark comes in action to deliver requested work. In this paper we are considering spark standalone cluster consists of five worker node including one of them work as master node as well. Apache Spark is a cluster computing platform as stated in \cite{karau2015learning, belouch2018performance} designed to be as fast as near real time response rate, Spark extends the popular MapReduce model to efficiently support more types of computations, including interactive queries and stream processing. It's speed
is important in processing large datasets, as it means the difference between exploring data interactively and waiting
minutes or hours. One of the main features Spark offers for speed is the ability to run computations in memory, but
the system is also more efficient than MapReduce for complex applications running on disk. Apache Spark is also designed to
cover a wide range of workloads that previously required separate distributed systems, including batch applications,
iterative algorithms, interactive queries, and streaming. By supporting these workloads in the same engine, Spark
makes it easy and cost effective to combine different processing types, which is often necessary in production data analysis pipelines. In addition, Apache spark reduces the management burden of maintaining separate tools. Spark also supports collection of different programming languages including R,Python,Scala etc.
Refer to the following subsection for configuration guidelines for getting started with Spark.
\subsubsection{Spark Cluster Configuration} 
Spark cluster is a network of computers. A cluster consist of a master and n* slaves. (N is limited) \\
Master: Master is the leader of all servers that monitors how slaves are working. It divides the task and take care of rest.\\
Slaves: These are the computers that receive job from master node and perform job. They are responsible to process chunks of your massive datasets following the Map Reduce paradigm. A computer can be master and slave at the same time.\\
Spark offers/supports three types of cluster including standalone, Yarn and Mesos.
Standalone: Spark is responsible to manage it's own cluster. \\
Yarn: Spark use Hadoop's Yarn resource manager. \\
Mesos: Spark Apache's dedicated resource manager.

\subsection{Result and Notification}
In this stage our proposed framework suggest to generate notification to avoid future vulnerabilities and to mitigate from current attacks. On successful completion of learning tasks, final result file will be transferred and admin server will get notified for anomalies (for this we are using command line based interface to track all interaction). Figure-\ref{fig:SAD-F} presents a complete illustration of our proposed framework.

\section{SAD-F: Scientific Datasets} \label{datasets}
Network anomaly detection is the challenging task due to dynamic nature of netflow traffic. The paper focuses on the general analytics/techniques to detect anomalies in network. The study exploring following datasets
\begin{itemize}
    \item KDD Cup 99
    \item UNSW-NB15
\end{itemize}
All of the mentioned datasets are in .pcap format provided by sources \cite{KDD,UNSW}. Further following subsection describes the detailed discussion of listed dataset individually.
\subsection{Datasets for System Evaluation}
This study does not require any specific traffic data because In this paper we are proposing a general framework. But we recommend these Netflow datasets KDD-CUP 99, and UNSW-NB15. Even though KDD-CUP 99 is not use full anymore because it is lacking most of the new types of attacks but we took this dataset for testing and proof of concept of our deployed framework.
\begin{itemize}
    \item KDD Cup 99: Since 1999, KDD99 noticed to be the widely used dataset for evaluation of anomaly detection methods \cite{tavallaee2009detailed,revathi2013detailed,olusola2010analysis}. This dataset is prepared by \cite{tavallaee2009detailed} and is built based on the data captured in DARPA’98 IDS evaluation
    program \cite{tavallaee2009detailed}. \\
    "This is the data set used for The Third International Knowledge Discovery and Data Mining Tools Competition, which was held in conjunction with KDD-99 The Fifth International Conference on Knowledge Discovery and Data Mining. The competition task was to build a network intrusion detector, a predictive model capable of distinguishing between ``bad'' connections, called intrusions or attacks, and ``good'' normal connections. This database contains a standard set of data to be audited, which includes a wide variety of intrusions simulated in a military network environment." The dataset's attacks fall in to four categories:
    \begin{itemize}
        \item Denial of Service Attack (DoS): DoS is one the attack type in which the intruder makes some computing or memory resource too busy to handle legitimate requests, or denies legitimate users access to a machine.
        \item User to Root Attack (U2R): It is a class of exploit in which the attacker starts out with access to a normal user account on the system (perhaps gained by sniffing passwords, a dictionary attack, or social engineering) and is able to exploit some vulnerability to gain root access to the system.
        \item Remote to Local Attack (R2L): It occurs when an attacker who has the ability to send packets to a machine over a network but who does not have an account on that machine exploits some vulnerability to gain local access as a user of that machine.
        \item Probing Attack: Probing is an attempt to gather information about a network of computers for the apparent purpose of circumventing its security controls.
    \end{itemize}
    \item UNSW-NB15: 
    For the evaluation of performance and effectiveness of network intrusion detection system, we require a comprehensive dataset which contains both normal and abnormal behaviors. Lot of research has been done using older benchmark data sets like KDDCUP 99 and NSLKDD but these data sets do not offer realistic output performance. The reason is that KDD CUP 99 has lots of redundant and missing records in the training set. So these datasets are not comprehensive representation of modern low foot print attack environment which concludes that we need a dataset which fulfills the requirements.
    
    The UNSW-NB 15 data set was created by utilizing an "IXIA PerfectStorm tool" to extract a hybrid of modern normal and contemporary attack activities of network traffic. A tcpdump tool was used to capture "100 GB" of raw network traffic (pcap files). Each pcap file contains "1000 MB" in order to make analysis of packets easier.
    
    UNSW-NB15 dataset is available in comma-separated values(CSV) file format.This data set contains 2,540,044 records which are stored in four CSV files. Moreover, a part from this data set was divided into a training set and a testing set. The training set involved 175,341 records, while the testing set contained 82,332 records with all different 9 types of attack and normal records. There are 45-49 attributes or features with 10 class values in this dataset. All records are divided in two major categories of the records - normal and attack. Furthermore the attack category is again subdivided into 9 categories of attack types. \cite{moustafa2015significant}. 
    \begin{itemize}
        \item Fuzzers: It is an attack in which the attacker attempts to discover security loopholes in an application, operating system or a network by feeding it with the massive inputting of random data to make it crash.
        \item Worms: It is an attack in which the attacker replicates itself to spread on other computers. Often, it utilizes a computer network to spread itself, depending on the security failures of the target computer used to access it.
        \item Reconnaissance: This attack category also known as probe, and it is an attack which gathers information about a computer network to evade its security controls.
        \item Analysis: It is a type of variety intrusions that penetrate the web applications via ports (e.g. port scans), emails (e.g. spam) and web scripts (e.g. HTML files).
        \item Backdoors: It is a technique of bypassing a stealthy normal authentication, securing unauthorized remote access to a device, locating the entrance to plain text, as it struggles to continue unobserved.
        \item DoS: It is an intrusion which disrupts the computer resources via memory so as to cause excessive business, in order to prevent authorized requests from accessing a device.
        \item Exploits: It is a sequence of instructions that takes advantage of a glitch, bug or vulnerability, causing an unintentional or unsuspected behavior on a host or a network.
        \item Generic: It is a technique that establishes against every block-cipher using a hash function to cause a collision without respect to the configuration of the block-cipher.
        \item Shellcode:It is a malware in which the attacker penetrates a slight piece of code starting from a shell to control the compromised machine.
    \end{itemize}
Furthermore, research studies \cite{moustafa2015significant,moustafa2015unsw} divided this dataset in to six different categories as follows:
\begin{itemize}
    \item Flow features: This group includes the identifier attributes between hosts, such as client-to-serve or server-to-client.
    \item Basic features: this category involves the attributes that represent protocols connections.
    \item Content features: this group encapsulates the attributes of TCP/IP; also they contain some attributes of http services.
    \item Time features: this category contains the attributes of time, for example, arrival time between packets, start/end packet time and round trip time of TCP protocol.
    \item Additional generated features: it includes general purpose features and connection features.
    \item Labelled Features: this group represents the label of each record.
\end{itemize}

\begin{table}[h!]
\begin{tabular}{|p{0.5cm}|p{1cm}|p{6cm}|} 
\hline
SNo & Feature Name & Feature Description \\  
\hline 
1 &srcip &Source IP address.\\
2 &sport &Source port number.\\
3 &dstip &Destinations IP address. \\
4 &dsport &Destination port number. \\
5 &proto &Protocol type, such as TCP, UDP. \\
\hline
\end{tabular}
\caption{Flow Features; {Table reproduced from scientific papers \cite{UNSW} and \cite{moustafa2015unsw,moustafa2015significant}}}
\label{table:flow}
\end{table}

\begin{table}[h!]
\begin{tabular}{|p{0.5cm}|p{1cm}|p{6cm}|}
 \hline
 SNo & Feature Name & Feature Description \\
 \hline 
6 &state &The states and its dependent protocol e.g., CON.\\
7 &dur & total duration.\\
8 &sbytes &Source to destination bytes.\\
9 &dbytes &Destination to source bytes.\\
10 &sttl &Source to destination time to live.\\
11 &dttl &Destination to source time to live.\\
12 &sloss &Source packets retransmitted or dropped.\\
13 &dloss &Destination packets retransmitted or dropped.\\
14 &service &Such as http, ftp, smtp, ssh, dns and ftpdata.\\
15 &sload &Source bits per second.\\
16 &dload &Destination bits per second.\\
17 &spkts &Source to destination packet count.\\
18 &dpkts &Destination to source packet count. \\ 
 \hline
\end{tabular}
\caption{Basic Features; {Table reproduced from scientific papers \cite{UNSW} and \cite{moustafa2015unsw,moustafa2015significant}}}
\label{table:basic}
\end{table}
\begin{table}[h!]
\begin{tabular}{|p{0.5cm}|p{1cm}|p{6cm}|}
 \hline
 SNo & Feature Name & Feature Description \\ 
 \hline 
19 &swin &Source TCP window advertisement value.\\
20 &dwin &Destination TCP window advertisement value.\\
21 &Stcpb &Source TCP base sequence number.\\
22 &dtcpb &Destination TCP base sequence number.\\
23 &smeansz &Mean of the packet size transmitted by the srcip.\\
24 &dmeansz &Mean of the packet size transmitted by the dstip.\\
25 &transdepth &The connection of http request or response transaction.\\
26 &resbdylen &The content size of the data transferred from http \\
 \hline
\end{tabular}
\caption{Content Features;{Table reproduced from scientific papers \cite{UNSW} and \cite{moustafa2015unsw,moustafa2015significant}}}
\label{table:content}
\end{table}
\begin{table}[h!]
\begin{tabular}{|p{0.5cm}|p{1cm}|p{6cm}|}
 \hline
 SNo & Feature Name & Feature Description \\
 \hline 
27 &sjit &Source jitter.\\
28 &djit &Destination jitter.\\
29 &stime &start time.\\
30 &ltime &last time.\\
31 &sintpkt &Source inter-packet arrival time.\\
32 &dintpkt &Destination inter-packet arrival time.\\
33 &tcprtt &Setup round-trip time, the sum of ’synack’ and ’ackdat’.\\
34 &synack &The time between the SYN and the SYN-ACK packets.\\
35 &ackdat &The time between the SYN-ACK and the ACK packets.\\
36 &ismipports &If srcip (1) = dstip (3) and sport (2) = dsport (4), assign 1 else 0. \\[1ex] 
 \hline
\end{tabular}
\caption{Time Features; {Table reproduced from scientific papers \cite{UNSW} and \cite{moustafa2015unsw,moustafa2015significant}}}
\label{table:time}
\end{table}

\begin{table}[h!]
\begin{tabular}{|p{0.5cm}|p{1.5cm}|p{5.5cm}|}
 \hline
SNo & Feature Name & Feature Description \\ [0.5ex] 
 \hline 
37 &ctstatettl &No. of each state (6) according to values of sttl (10) and dttl (11).\\
38 &ctflwhttpmthd &No. of methods such as Get and Post in http service.\\
39 &isftplogin &If the ftp session is accessed by user and password then 1 else 0.\\
40 &ctftpcmd &No of flows that has a command in ftp session.\\
41 &ctsrvsrc &No. of rows of the same service (14) and srcip (1) in 100 rows.\\
42 &ctsrvdst &No. of rows of the same service (14) and dstip (3) in 100 rows.\\
43 &ctdstltm &No. of rows of the same dstip (3) in 100 rows.\\
44 &ctsrcltm &No. of rows of the srcip (1) in 100 rows.\\
45 &ctsrcdportltm &No of rows of the same srcip (1) and the dsport (4) in 100 rows.\\
46 &ctdstsportltm &No of rows of the same dstip (3) and the sport (2) in 100 rows.\\
47 &ctdstsrcltm &No of rows of the same srcip (1) and the dstip (3) in 100 records.\\[1ex] 
 \hline
\end{tabular}
\caption{Additional generated Features;{Table reproduced from scientific papers \cite{UNSW} and \cite{moustafa2015unsw,moustafa2015significant}}}
\label{table:additional}
\end{table}
\begin{table}[h!]

\begin{tabular}{|p{0.5cm}|p{1cm}|p{6cm}|}
\hline
 SNo & Feature Name & Feature Description \\  
 \hline 
48 &Attackcat &The name of each attack category.\\
49 &Label &0 for normal and 1 for attack records. \\
 \hline
\end{tabular}
\caption{Labelled  Features;{Table reproduced from scientific papers \cite{UNSW} and \cite{moustafa2015unsw,moustafa2015significant}}}
\label{table:labelled}
\end{table}
\end{itemize}

From the selected datasets, we focus on a specific group of attacks as discussed above in subsection \ref{datasets}. The considered algorithms were trained to detect each of these attack types independently and in parallel, in the same fashion as in \cite{casas2017network,vanerio2017ensemble}. As a result, each detection approach can detect the occurrence of an attack and also classify its nature.

\section{Testbed Design Steps} \label{testbed}
This section highlights and discuss the implementation design steps of proposed framework. The following subsection are small components which need to configured before the execution of system.
\subsection{Testbed setup}
We design and tested our framework on two different testbed we named one as "low end" testbed and "high end" testbed. Both have different specification in terms of hardware.  

For low end testbed we configured 3 worker nodes and 1 master node. Each node is physically independent of each other. Each system contains 8GB of memory and 500 GB of Hard Disk and 2.7 GHz of processor. Collectively this cluster setup consist of 10*16 GB of memory, 10*500 GB of space and 3*3.5 GHz of processor as per configuration of previous study \cite{casas2017network}.
    
For high end testbed we configured 2 worker nodes and one master node. Each node is physically independent of each other. Each system contains 16GB of memory and 1TB of Hard Disk and 3.5 GHz of processor. Collectively this cluster setup consist of 2*16 GB of memory, 2*1 TB of space and 3*3.5 GHz of processor.
    
\subsection{Netflow traffic selector}
As shown in figure-\ref{fig:SAD-F}, our proposed framework is flexible with either live traffic or passive traffic selector. At the current stage we only used passive traffic selector. At this step system requires required traffic file after that system will transfer that file to the log file for further processing. Following two option are available as traffic selector.
    \begin{itemize}
        \item Live network traffic capturing server
        \item Passive traffic
    \end{itemize}
\subsection{Log server} After receiving file at this stage log server will kept that file for future need if there is any discrepancy in results admin/network administrator can do manual work to take respective actions.
\subsection{Preprocessing file}We designed a preprocessing function which is efficient and it is designed with respect to the netflow dataset but it is quite easy to change it's parameters to fit with a new netflow dataset, following are the few steps which we considered while designing preprocessing module.
    \begin{itemize}
        \item Data file splitting by time stamp
        \item Calculating hash of features at run-time
        \item Converting features into OneHotEncoding at run-time
    \end{itemize}
    Before beginning for further classification of network traffic, we split targeted data by time-slot to reduce the size of file. For example, a one hour  network trace file were split in multiple 10-Seconds files. General command to split network trace file based on time-slot. "editcap -i file-split-time filename.pcap file-split-time.pcap"
    
    For preprocessing, we designed our own methodology which aims to read the data file and load it into data-frame. Then data-frame were read individually by features set. By Deep feature inspection we noticed data type of each feature after that we used appropriate conversion method for each feature refer to tables [\ref{table:flow}, \ref{table:basic}, \ref{table:content}, \ref{table:time}, \ref{table:additional} and \ref{table:labelled}] for detailed dataset feature set.
\subsection{File transfer to spark cluster}After taking all the necessary steps, preprocessed file arrives at spark phase on which system distributes the file to the worker node which is one of the spark's core job.
\subsection{Model selection}We implemented five supervised learning algorithms {KNN, RF, DT, NB and SVM} and one unsupervised algorithm {kMeans} to train as referenced in \cite{casas2017network}.
\subsection{Model testing} We used training model as our base model to classify whether the incoming network traffic to the server or main node is normal or an attack.
\subsection{Admin panel}Whoever will deploy our framework to improve their network operation in terms of DDoS detection they can design their admin panel as per requirements. The framework is capable of further integration. At this stage we used Ubuntu command prompt system to tackle admin related tasks like start/stop the operations.
\section{Experiments and Result Analysis} \label{experiments}
The framework is independent of netflow dataset selection, however we explored and practically implemented the UNSW-NB-15 dataset is publicly available at \cite{UNSW} since 2015 and to the best of our knowledge this is the most suitable and well described dataset available for researchers to work.
\textbf{Apache Spark Framework:} Apache Spark based anomaly detection framework and MLib libraries are used. Five well-known machine learning algorithms are used, namely [KNearestNeighbour, NaïveBayes, DecisionTree, SuportVectorMachine and RandomForest] are used for performance evaluation. Recently, similar work has been proposed by \cite{belouch2018performance}. One may compare the significant result difference of our proposed work with this work.

\subsection{Performance evaluation:} The overall performance of SAD-F depends upon time taken to capture and transfer log file to the testbed in real time mode. The system is totally depend upon time taken to read chunk of traffic volume and time taken to classify each chunk and moves forward for next chunk of traffic as in upcoming tables [1-3] we presented two variation of chunk size i-e 300 and 1000  and it also depend upon the time taken by specific classifier to detect chunk and generate false alarm + negligible overhead of the system.

For performance evaluations we benchmark our results which focuses on selection of classifier, file size, total volume of traffic, false alarm rate, chunk size of traffic volume and detection time.

Figure-\ref{fig:my_label} depicts total time taken to capture and transfer real traffic (using Wireshark T-Shark library) on our deployed main node. Traffic capturing time is almost linear to the file size, as the file size is increasing it is also increasing. It took almost 20 seconds to capture a file size of 10MB while it took 420 seconds to capture 1 GB file on the other hand it may also noticed that as file size is increasing the volume of traffic is also increasing so both relations are linear co-related. This shows a clear improvement in throughput with the increase in file size and volume of traffic. 

One point must be noted at this stage that these calculations are the proof of concept while we did not continued with real time strategy although we are reading chunk of data and simulating like a real time (near real time) but these measurements will be use full for future researcher. 

\begin{figure}
    \centering
    \includegraphics[width=0.9\linewidth]{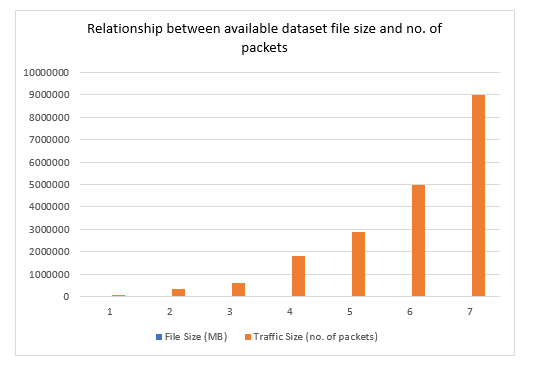}
    \caption{Relationship of Dataset Size and Traffic Volume}
    \label{fig:datasetSize}
\end{figure}

Figure-\ref{fig:datasetSize} list dataset volume w.rt. to file size  and packets count.

\begin{figure}
    \centering
    \includegraphics[width=0.9\linewidth]{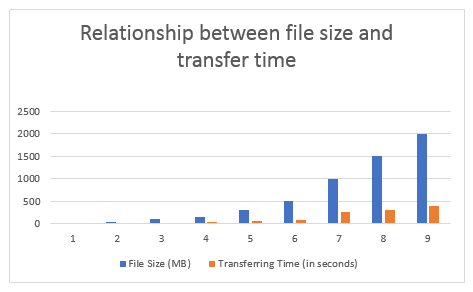}
    \caption{Capture and transfer time of a log file.}
    \label{fig:my_label}
\end{figure}

\subsubsection{Comparison of preprocessing, training and testing cost on local machine vs SAD-Framework}
Scientific reading presented in this Table \ref{table:localmachine} taken from local machine (8GB of RAM and 2.50 GHz of processor) while reading presented in Table \ref{table:lowend} and \ref{table:highend} are taken from SAD-Framework.
\begin{table}[h]
\centering
\begin{tabular}{|p{1cm}|p{3cm}|p{1cm}|p{1cm}|} 
 \hline
Model & Train/Test file size and Traffic volume & Overall Time & Accuracy \\ 
 \hline 
    kNN	& 150 MB 40 MB (999999, 300000) & 12605 & 92 \\ 
    DT	& 150 MB 40 MB (999999, 300000)	& 6565 & 94.40 \\
    RF	& 150 MB 40 MB (999999, 300000)	& 6549 & 92.98 \\
    NB	& 150 MB 40 MB (999999, 300000)	& 6536 & 72.6 \\
    SVM& 150 MB  40 MB (500000, 200000)	& 71317& 87.62 \\ [1ex]
 \hline
\end{tabular}
\caption{Relationship of train/test time (in seconds) and Accuracy of model in percentage on a local machine}
\label{table:localmachine}
\end{table}

\begin{table}[h]
\centering
\begin{tabular}{|p{1cm}|p{3cm}|p{1cm}|p{1cm}|} 
 \hline
 Model & Train/Test file size and Traffic volume & Overall Time & Accuracy \\ [0.5ex] 
 \hline\hline
    DT	& 150 MB 40 MB (999999, 300000)	& 90 & 94.40 \\
    RF	& 150 MB 40 MB (999999, 300000)	& 360 & 92.98 \\
    NB	& 150 MB 40 MB (999999, 300000)	& 204 & 72.6 \\
    SVM & 150 MB  40 MB (500000, 200000)& 14400 & 87.62 \\ [1ex]
 \hline
\end{tabular}
\caption{Relationship of train/test time (in seconds) and Accuracy of model in percentage on a low end testbed}
\label{table:lowend}
\end{table}
\begin{table}[h]
\centering
\begin{tabular}{|p{1cm}|p{3cm}|p{1cm}|p{1cm}|} 
 \hline \sloppy
 Model & Train/Test file size and Traffic volume & Overall Time & Accuracy \\ [0.5ex] 
 \hline\hline
    DT	& 150 MB 40 MB (999999, 300000)	& 32 & 92.23 \\
    RF	& 150 MB 40 MB (999999, 300000)	& 200 & 91.68 \\
    NB	& 150 MB 40 MB (999999, 300000)	& 104 & 78.54 \\
    SVM & 150 MB  40 MB (500000, 200000)& 2890 & 86.52 \\ [1ex]
 \hline
\end{tabular}
\caption{Relationship of train/test time (in seconds) and Accuracy of model in percentage on a high end testbed}
\label{table:highend}
\end{table}
These reading were taken from low end testbed. Measurements of KNN is not presented in above tables \ref{table:lowend} and \ref{table:highend} because kNN is not supported by SPARK distributed architecture \cite{sparkKNN}. 

By analysing readings from above presented tables \ref{table:localmachine}, \ref{table:lowend} and \ref{table:highend} which clearly shows the improvement. Here we conclude our proof of work that our deployed frameworks. Next section focuses on actual scientific reading taken from SAD-F on both testbed individually by simulating near real time traffic.

\subsection{Near real time traffic classification by SAD-F:} Our system has the capability to simulate traffic in form of single/multiple rows and in network’s terminology chunk of traffic volume after that our program will pass this traffic to classifier/function to classify the traffic behavior and generate false alarm if system detected as any malware available in selected chunk of data and repeats the process to select another chunk of volume till end of file. 

This simulated traffic can also be a real time traffic refer to figure-\ref{fig:my_label}, but it needs an extra amount of efforts in terms of time for preprocessing to include all the required features set for proper working and to increase accuracy and time taken by logging and transferring such traffic volume to the testbed/deployed framework.

We evaluated the performance of our framework on the basis of different size of the log file which is depicted in figure-\ref{fig:datasetSize}. For testbed evaluation we individually noted spark file loading and distributing time, file preprocessing time, detection time of algorithm and total time taken by spark cluster to complete the experiment. For the scientific reason we calculated average of three measurements for individual experiments. Following paragraphs shows low end testbed and high end testbed results respectively. 
\subsection{Low end Testbed Result Charts}
\begin{table}[h]
\centering
\begin{tabular}{|c c c c c|} 
 \hline
Packet Count &DT	&RF	&NB	&SVM
 \\ [0.5ex] 
 \hline 
100K	&1.2008	&0.9705	&0.7536	&153.2593 \\
450K	&1.5635	&1.5776	&5.9583	&543.8486\\
1000K	&3.3746	&2.8163	&21.0618	&946.4972\\
1800K	&179.453	&66.4375	&38.9206	&1176.6666\\
2500K	&428.3554	&376.1895	&442.9408	&2008.8368\\
3400K	&643.8411	&613.1385	&850.6137&2900.5085\\[1ex] 
 \hline
\end{tabular}
\caption{Detection time taken by different classifier w.r.t Packet Size}
\label{table:DT}
\end{table}
\begin{figure}[h!]
    \centering
    \includegraphics[width=0.9\linewidth]{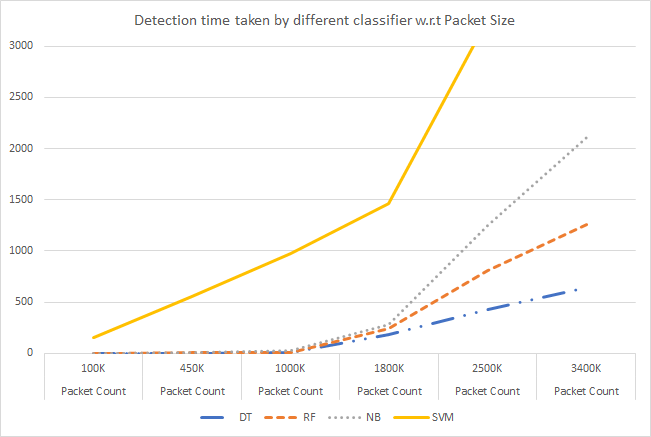}
    \caption{Detection time taken by different classifier w.r.t Packet Size}
    \label{fig:DT}
\end{figure}

Table \ref{table:DT} shows the average detection time taken by different detection models namely [DT: Decision Tree, RF: Random Forest, NB: Naive Bayes and SVM: Support Vector Machine]. We noticed that NB performed better than all three remaining models and took 0.6 and 2.5 seconds for 100K and 450K packets count respectively and RF took 4.46 seconds for 1000K packets count. We also noticed that on average RF performed well w.r.t all three models and took maximum 335 seconds for 3400K packets count while SVM is taking maximum time w.r.t all three models. SVM took almost 90 seconds for 10MB log file which consist of 100K packets count and it took approx 860 seconds for 500 MB file which reflects 3400K packet count. We also presented these results in pictorial form which is depicted in figure-\ref{fig:DT}.

Figure figure-\ref{fig:DT} shows complete picture of detection time taken by different detection models. On horizontal axis we placed total packet counts I.e [100K, 450K, 1000K, 1800K, 2500K and 3400K]. On vertical axis we placed time in seconds. This figure-\ref{fig:DT} is exact representation of the above \ref{table:DT}. 

\begin{table}[h!]
\centering
\begin{tabular}{|c c c c c|} 
 \hline
File Size &DT	&RF	&NB	&SVM
 \\ [0.5ex] 
 \hline 
10MB	&0.3833	&0.204	&0.4069	&1.0088\\
50MB	&4.1349	&0.5576	&1.7388	&1.2349\\
100MB	&5.4611	&0.5618	&2.6002	&3.2608\\
200MB	&8.8026	&0.5812	&4.5489	&9.1338\\
400MB	&53.7333	&2.9055	&32.6329	&15.5418\\
500MB	&98.2872	&60.3406	&121.6788	&26.4873 \\[1ex]
 \hline
\end{tabular}
\caption{File loading and distributing time taken by spark w.r.t file size}
\label{table:FILETIME}
\end{table}
\begin{figure}[h!]
    \centering
    \includegraphics[width=0.9\linewidth]{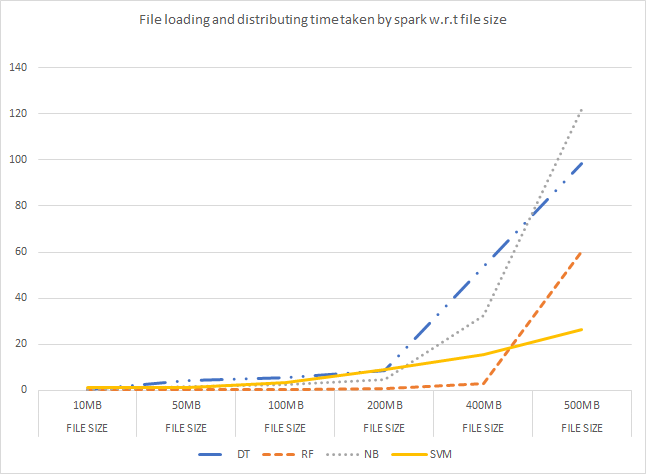}
    \caption{File loading and distributing time taken by spark w.r.t file size}
    \label{fig:FILETIME}
\end{figure}

Table \ref{table:FILETIME} shows the average file loading and distribution time taken by spark cluster w.r.t to file size in mega bytes. Reading shows that least time taken by spark 0.204 seconds for 10MB file while the maximum time taken by spark for the file size of 500MB is 98 seconds that is almost 1.5 minutes. 

We also presented the above tubular form results in the pictorial form below refer to the figure figure-\ref{fig:FILETIME}. It shows complete picture of file time taken by different models and log file size .On horizontal axis we placed log file size in mega bytes I.e [10M, 50MB, 100M, 400MB and 500MB]. On vertical axis we placed time in seconds I.e vary from 1 seconds to 120 seconds.
\begin{table}[h!]
\centering
\begin{tabular}{|c c c c c|} 
 \hline
Packet Count &DT	&RF	&NB	&SVM
 \\ [0.5ex] 
 \hline 
100K	&1.5676	&1.2247	&0.5849	&89.9029\\
450K	&3.4877	&2.5367	&2.5141	&137.0856\\
1000K	&7.7287	&4.4672	&4.4725	&281.1723\\
1800K	&25.3764	&9.1906	&8.7517	&642.9673\\
2500K	&180.3885	&216.0378	&94.5724	&724.8311\\
3400K	&295.8082	&335.2831	&338.404	&861.5172\\[1ex]
 \hline
\end{tabular}
\caption{Preprocessing time taken by model w.r.t file size}
\label{table:prepTime}
\end{table}
\begin{figure} [h!]
    \centering
    \includegraphics[width=0.9\linewidth]{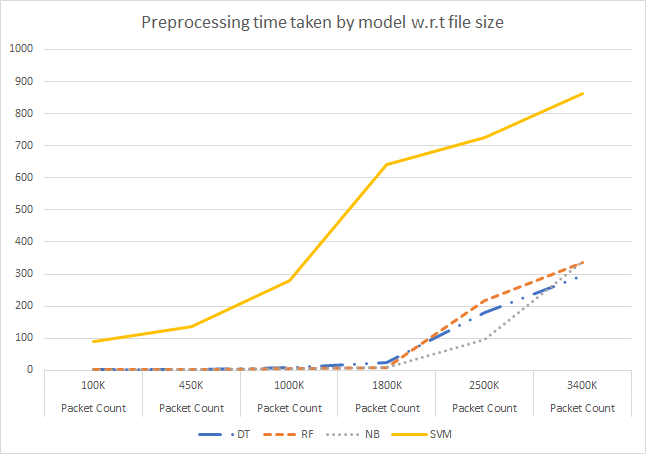}
    \caption{Preprocessing time taken by model w.r.t file size}
    \label{fig:prepTime}
\end{figure}
Table \ref{table:prepTime} and figure-\ref{fig:prepTime} respectively shows the average of three scientific reading of time taken by each models with respect to the total packets count. We noticed that preprocessing time increases as the file size or packets counts increases in every case but at the same time we noticed that it is not directly proportional to packets count but it is only dependent upon the machine learning model how fast it preprocess the small chunk of packet count and how much time it will take for large chunk of data. 
Naive Bayes remains the best preprocessing model for 100K packets and it took only 0.6 seconds to preprocess while at the same time it took approx 338 seconds for 3400K packets count. On average RF were remain the best model among all it took only [1.23 seconds and 335 seconds] for 100K and 3400K packets counts respectively. SVM remains the worst pre processor as well as worst classifier refer to table \ref{table:DT} and figure-\ref{fig:DT}.

\begin{table} [h!]
\centering
\begin{tabular}{|c c c c c|} 
 \hline
File Size &DT	&RF	&NB	&SVM \\ [0.5ex] 
 \hline 
10MB	&6.3333	&5.0001	&4.1666	&352.6666\\
50MB	&11.0002	&7.3333	&11.3333	&706.6666\\
100MB	&19.0001	&11.3333	&36.0001	&1270.3333\\
200MB	&252.6666	&105.6666	&165.6666	&2086.6666\\
400MB	&707.3333	&644.3333	&614.3333	&3466.6666\\
500MB	&1079.3333	&1075.6666	&1550.6666	&4460.3333 \\[1ex]
 \hline
\end{tabular}
\caption{Time taken by spark w.r.t file size a.k.a SPARK UI TIME}
\label{table:sparkTime}
\end{table}
\begin{figure}[h!]
    \centering
    \includegraphics[width=0.9\linewidth]{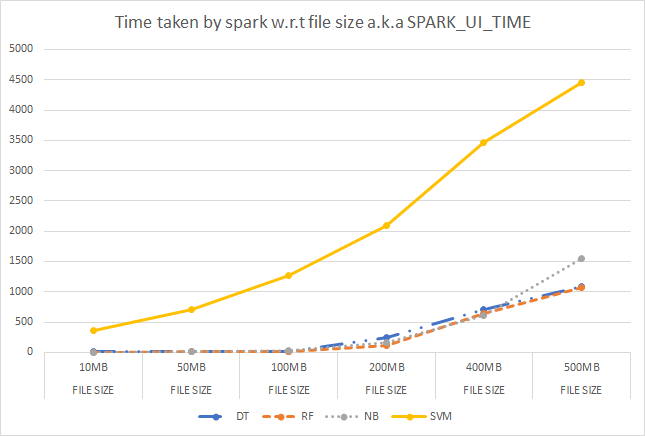}
    \caption{Time taken by spark w.r.t file size a.k.a SPARK UI TIME}
    \label{fig:sparkTime}
\end{figure}
Table \ref{table:sparkTime} lists the complete time taken by spark user interface which is actual sum of all individual times including (file time, preprocessing time and detection/classification time).figure-\ref{fig:sparkTime} is the pictorial presentation of above mentioned table.
\begin{figure} [h!]
    \centering
    \includegraphics[width=0.9\linewidth]{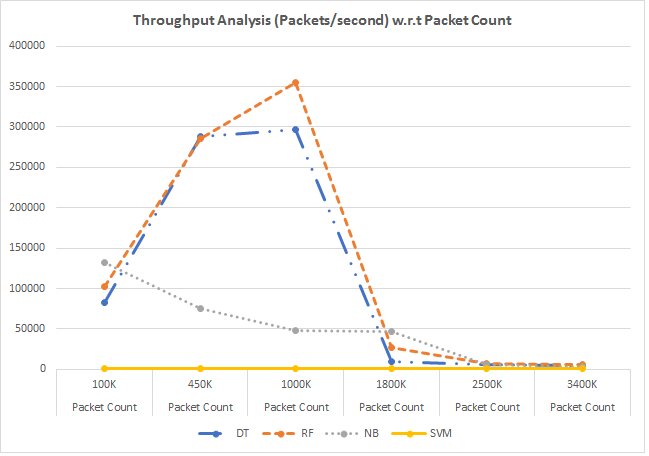}
    \caption{Throughput Analysis (Packets/second) w.r.t Packet Count}
    \label{fig:throughput}
\end{figure}
\begin{figure}[h!]
    \centering
    \includegraphics[width=0.9\linewidth]{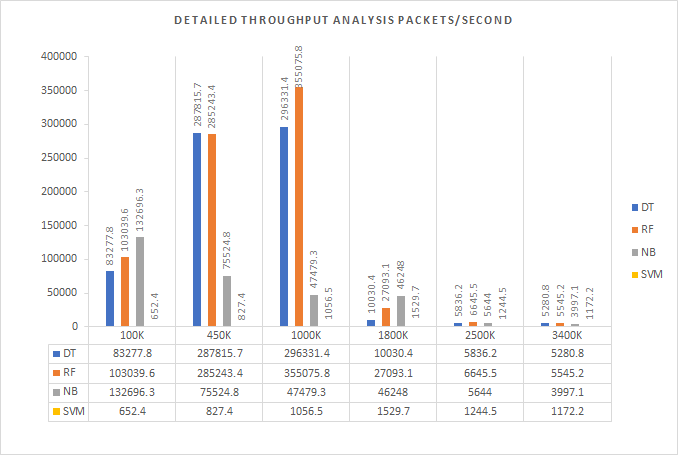}
    \caption{Detailed throughput analysis packets/second}
    \label{fig:DetailedTroughput}
\end{figure}
Refer to Figure[\ref{fig:DT}, \ref{fig:FILETIME}, \ref{fig:prepTime}, \ref{fig:sparkTime}, \ref{fig:throughput} and \ref{fig:DetailedTroughput}] for low testbed results conclusion. figure-\ref{fig:DT} and Table \ref{table:DT} shows detection time taken by specific model/classifier to classify packets with respect to the packet count. We noticed that for both [Decision Tree and Random Forest] models took less 5 seconds to classify 100K packets while [Random Forest] model took approx 21 seconds and SVM took maximum time than all three. We also noticed that overall [Random Forest] perform better than all other classifier in terms of detection packets counts per second.
\subsection{High end Testbed Result Charts}
\begin{table} [h!]
\centering
\begin{tabular}{|c c c c c|} 
 \hline
Packet Count &DT	&RF	&NB	&SVM \\ [0.5ex] 
 \hline 
100K	&0.1707	&0.4046	&2.518	&305.1077\\
450K	&0.5454	&1.2576	&7.1007	&454.2456\\
1000K	&1.8796	&3.0908	&11.2807	&916.556\\
1800K	&5.0688	&5.5479	&27.5153	&1540.5229\\
2500K	&6.6027	&93.9212	&180.1262	&1718.1377\\
3400K	&79.8791	&156.749	&478.5542	&1844.2807 \\[1ex]
 \hline
\end{tabular}
\caption{Detection time taken by different classifier w.r.t Packet Size}
\label{table:DT1}
\end{table}
\begin{figure} [h!]
    \centering
    \includegraphics[width=0.9\linewidth]{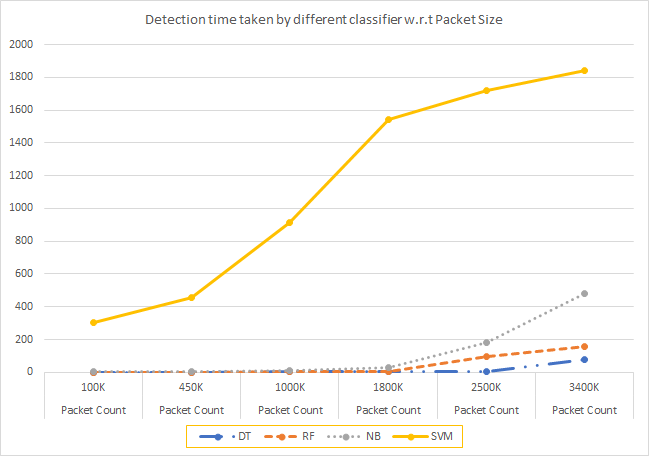}
    \caption{Detection time taken by different classifier w.r.t Packet Size}
    \label{fig:DT1}
\end{figure}
Table \ref{table:DT1} shows the average detection time taken by different detection models namely [DT: Decision Tree, RF: Random Forest, NB: Naive Bayes and SVM: Support Vector Machine]. We noticed that DT performed better than all three remaining models and took 0.17 and 0.5 seconds for 100K and 450K packets count respectively while DT took only 80 seconds to process 3400K packets. Which is far better than our previous spark cluster i.e. low end testbed. On other hand RF took 0.4 seconds for 100K packets count. We also noticed that on average RF performed well w.r.t NB and SVM while SVM is taking maximum time w.r.t all three models. SVM took almost 305 seconds for 10MB log file which consist of 100K packets count and it took approx 1844 seconds for 500 MB file which reflects 3400K packet count. We also presented these results in pictorial form which is depicted in Figure- \ref{fig:DT1}.

figure-\ref{fig:DT1} shows complete picture of detection time taken by different detection models. On horizontal axis we placed total packet counts I.e [100K, 450K, 1000K, 1800K, 2500K and 3400K]. On vertical axis we placed time in seconds. This figure-\ref{fig:DT1} is exact representation of the above \ref{table:DT1}.
\begin{table} [h!]
\centering
\begin{tabular}{|c c c c c|} 
 \hline
File Size &DT	&RF	&NB	&SVM \\ [0.5ex] 
 \hline 
10MB	&0.4007	&0.2865	&0.4239	&0.5681\\
50MB	&1.0771	&0.2757	&1.0425	&0.7659\\
100MB	&1.7559	&0.2793	&1.7353	&1.7181\\
200MB	&4.2826	&0.2687	&3.4016	&3.1394\\
400MB	&5.305	&0.272	&7.3116	&9.2446\\
500MB	&11.4324	&0.2751	&9.8913	&11.9497 \\[1ex]
 \hline
\end{tabular}
\caption{File loading and distributing time taken by spark w.r.t file size}
\label{table:FILETIME1}
\end{table}
\begin{figure}
    \centering
    \includegraphics[width=0.9\linewidth]{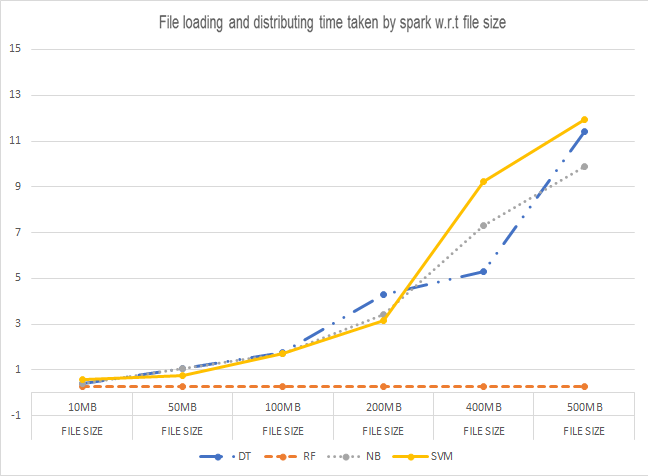}
    \caption{File loading and distributing time taken by spark w.r.t file size}
    \label{fig:FILETIME1}
\end{figure}

Table \ref{table:FILETIME1} presents the average file loading and distribution time taken by spark cluster w.r.t to file size in mega bytes. Reading shows that least time taken by spark 0.2865 seconds for 10MB file while the maximum time taken by spark for the file size of 500MB is approx 12 seconds which is also better than our low end cluster.

We also presented the above tubular form results in the pictorial form below refer to the figure-\ref{fig:FILETIME1}. It shows complete picture of file time taken by different models and log file size .On horizontal axis we placed log file size in mega bytes I.e [10M, 50MB, 100M, 400MB and 500MB]. On vertical axis we placed time in seconds I.e vary from 1 seconds to 120 seconds.
\begin{table}[h]
\centering
\begin{tabular}{|c c c c c|} 
 \hline
Packet Count &DT	&RF	&NB	&SVM
 \\ [0.5ex] 
 \hline 
100K	&0.5234	&0.5074	&1.1769	&2.6776\\
450K	&1.5384	&1.433	&3.1147	&4.5669\\
1000K	&3.4672	&3.2363	&5.259	&6.524\\
1800K	&8.9085	&6.0414	&11.2844	&82.0879\\
2500K	&18.3818	&11.5615	&94.2082	&359.9464 \\
3400K	&20.7432	&19.8323	&122.1618	&564.5891 \\[1ex]
 \hline
\end{tabular}
\caption{Preprocessing time taken by model w.r.t file size}
\label{table:prepTime1}
\end{table}
\begin{figure}[h!]
    \centering
    \includegraphics[width=0.9\linewidth]{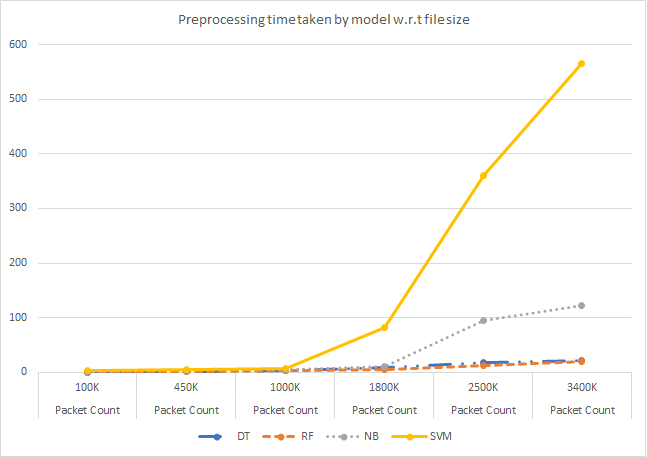}
    \caption{Preprocessing time taken by model w.r.t file size}
    \label{fig:prepTime1}
\end{figure}
Table \ref{table:prepTime1} and figure-\ref{fig:prepTime1} respectively shows the average of three scientific reading of time taken by each models with respect to the total packets count. We noticed that preprocessing time increases as the file size or packets counts increases in every case but at the same time we noticed that it is not directly proportional to packets count but it is only dependent upon the machine learning model how fast it preprocess the small chunk of packet count and how much time it will take for large chunk of data. 

Random Forest remains the overall best preprocessing model for 100K till 3400K packets and it took only 0.5 seconds to preprocess 100K while at the same time it took approx 19.8 seconds for 3400K packets count. While DT, NB performed well w.r.t low end testbed and SVM remains the worst pre processor as well as worst classifier refer to table \ref{table:DT1} and figure-\ref{fig:DT1}.

\begin{table}[h!]
\centering
\begin{tabular}{|c c c c c|} 
 \hline
File Size &DT	&RF	&NB	&SVM
 \\ [0.5ex] 
 \hline 
10MB	&4.3333	&5.3333	&6.0001	&311.6666\\
50MB	&6.0001	&7.0001	&12.0001	&522.3333\\
100MB	&8.0001	&9.3333	&19.0001	&965.6666\\
200MB	&19.3333	&16.6666	&44.6666	&1648\\
400MB	&32.3333	&126.3333	&289.3333	&2118.3333 \\
500MB	&191.0001	&201.3333	&647.3333	&2493 \\[1ex]
 \hline
\end{tabular}
\caption{Time taken by spark w.r.t file size a.k.a SPARK UI TIME}
\label{table:sparkTime1}
\end{table}

\begin{figure}[h!]
    \centering
    \includegraphics[width=0.9\linewidth]{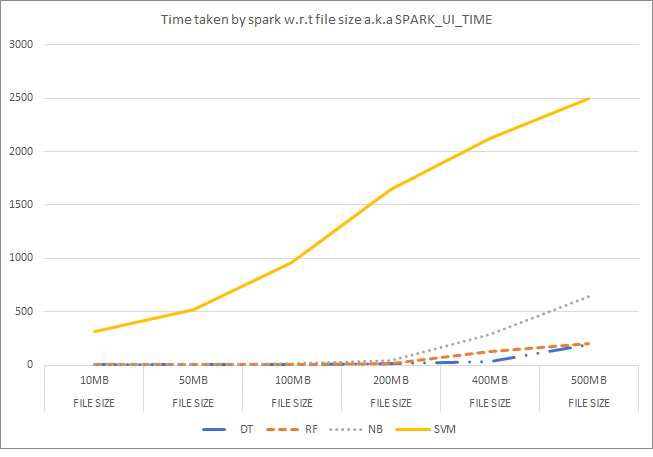}
    \caption{Time taken by spark w.r.t file size a.k.a SPARK UI TIME}
    \label{fig:sparkTime1}
\end{figure}

Table \ref{table:sparkTime1} lists the complete time taken by spark user interface which is actual sum of all individual times including (file time, preprocessing time and detection/classification time).figure-\ref{fig:sparkTime1} is the pictorial presentation of above mentioned table.

\begin{figure}[h!]
    \centering
    \includegraphics[width=0.9\linewidth]{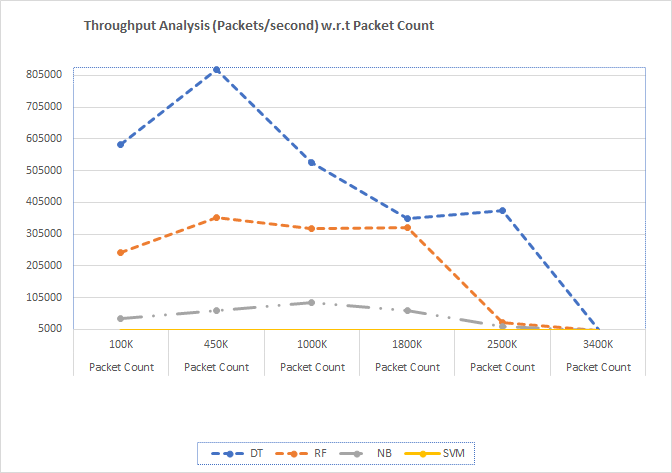}
    \caption{Throughput Analysis (Packets/second) w.r.t Packet Count}
    \label{fig:throughput1}
\end{figure}
\begin{figure}[h]
    \centering
    \includegraphics[width=0.9\linewidth]{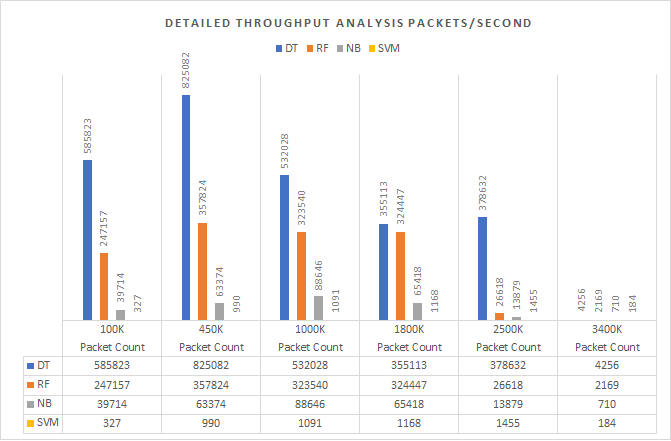}
    \caption{Detailed throughput analysis packets/second}
    \label{fig:DetailedTroughput1}
\end{figure}
Refer to figures[\ref{fig:DT1}, \ref{fig:FILETIME1}, \ref{fig:prepTime1}, \ref{fig:sparkTime1}, \ref{fig:throughput1} and \ref{fig:DetailedTroughput1}] for low testbed results conclusion. figure-\ref{fig:DT1} and Table \ref{table:DT1} shows detection time taken by specific model/classifier to classify packets with respect to the packet count. We noticed that for both [Decision Tree and Random Forest] models took less five seconds to classify 1800K packets while [Random Forest] model took approx 27 seconds and SVM took maximum time than all three. here reader can see the actual difference in time to detect number of packets with "low end" and "high end" testbed. We also noticed that overall [Random Forest] perform better than all other classifier in terms of detection packets counts per second. It took only 156 seconds to detect 3400K packets which is 400 percent better than low end.

\subsection{Overall discussion}
We computed and presented actual results in terms of detection time and remaining referenced figures[\ref{fig:FILETIME1}, \ref{fig:prepTime1} and \ref{fig:sparkTime1}], and tables[\ref{table:FILETIME1}, \ref{table:prepTime1} and \ref{table:sparkTime1}] are for reference to the reader for self understanding that how much time (in seconds) our framework took to load file, to preprocess and last but not the least how much actual time taken by spark testbed on each run.

Figures[\ref{fig:throughput} and \ref{fig:throughput1}] are representing throughput of both testbeds respectively. While figures[\ref{fig:DetailedTroughput} and \ref{fig:DetailedTroughput1}] are showing additional information of throughput and referenced both charts are additional efforts to understand results in better manner.

\begin{figure*}
    \centering
    \includegraphics[width=0.9\linewidth]{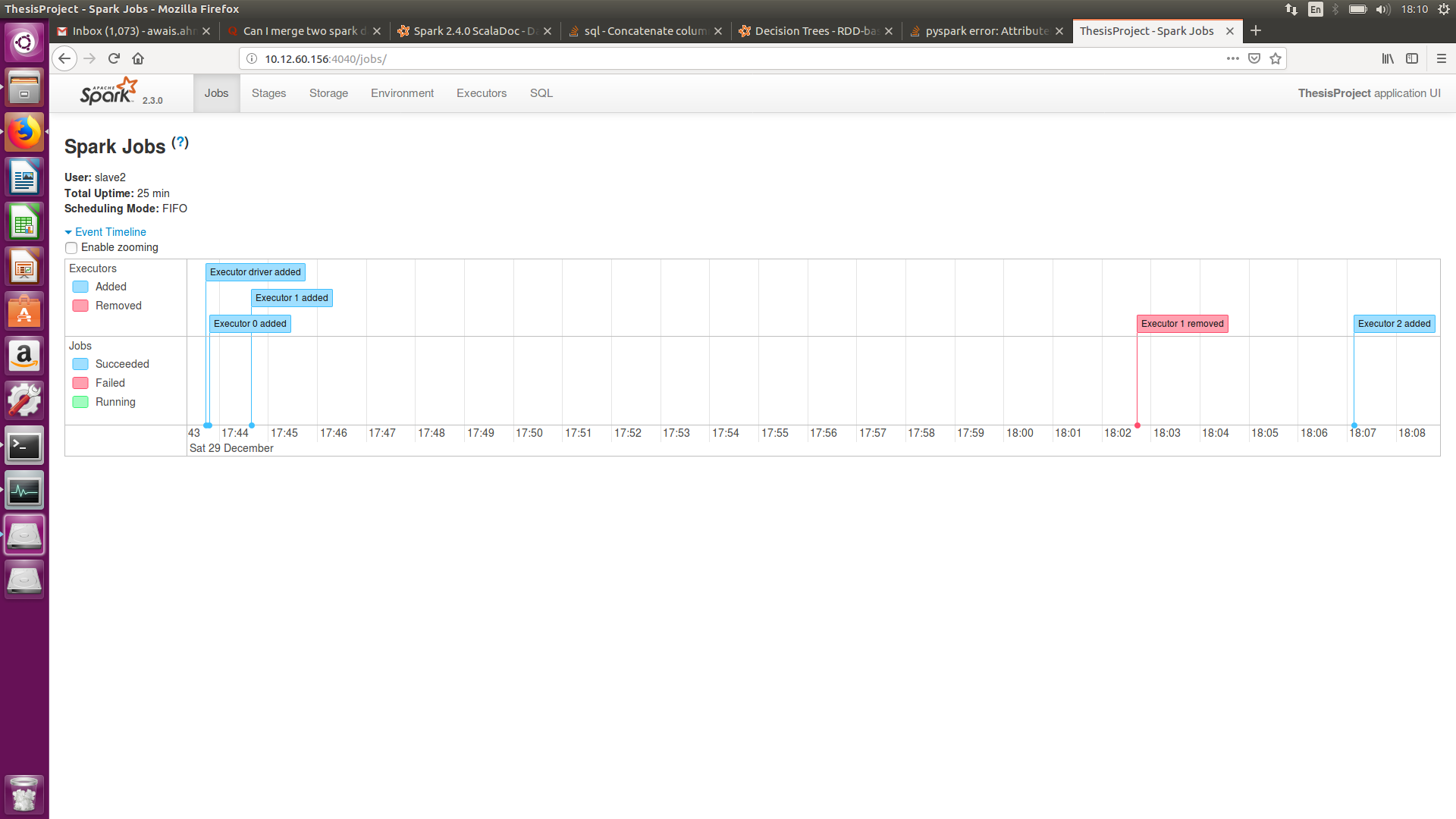}
    \caption{View of Spark Executor after submitting job to testbed}
    \label{fig:jobEx}
\end{figure*}

\begin{figure*}
    \centering
    \includegraphics[width=0.9\linewidth]{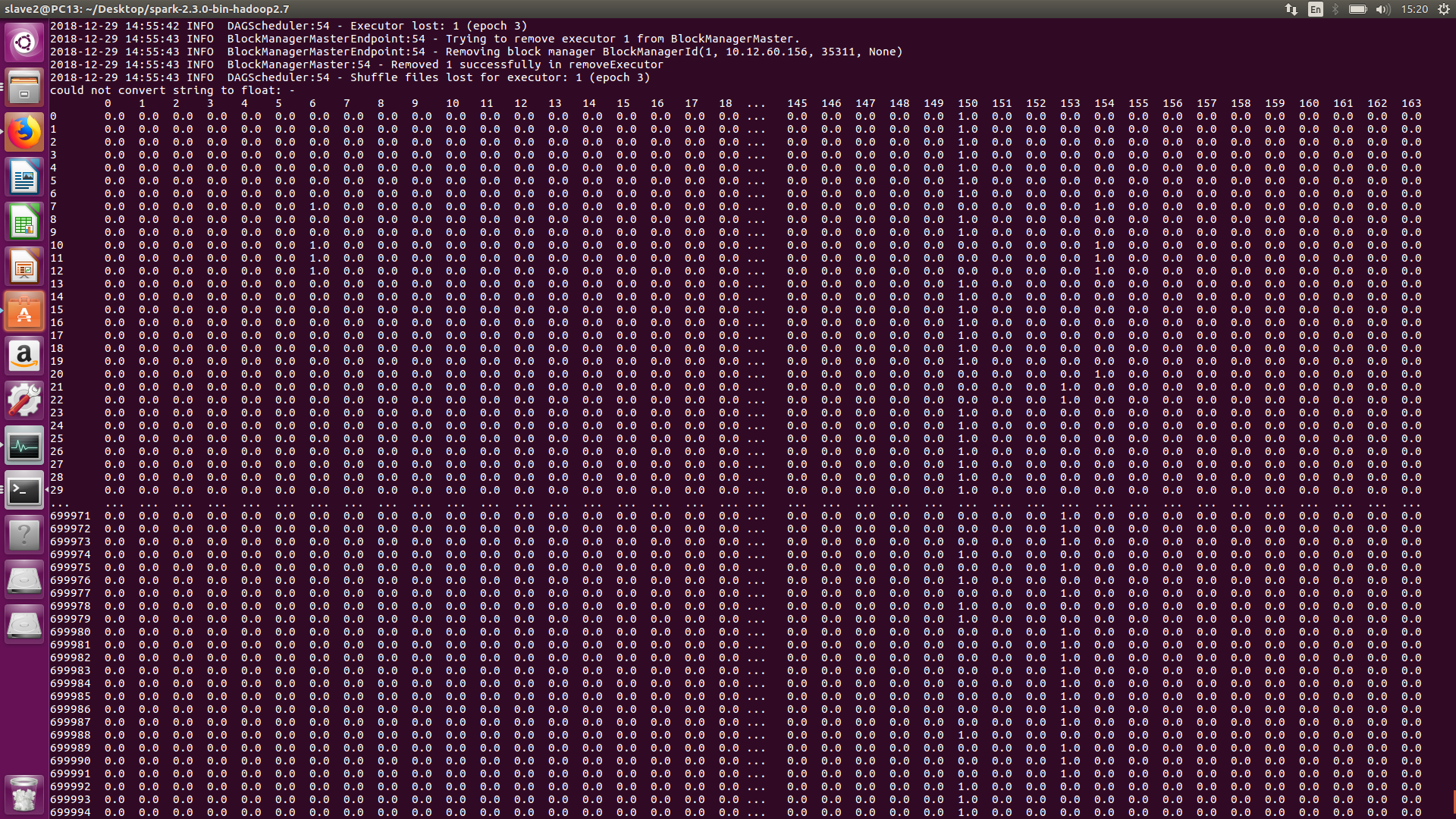}
    \caption{SAD-F: View of Preprocessing of Data}
    \label{fig:datPrep}
\end{figure*}

\begin{figure*}
    \centering
    \includegraphics[width=0.9\linewidth]{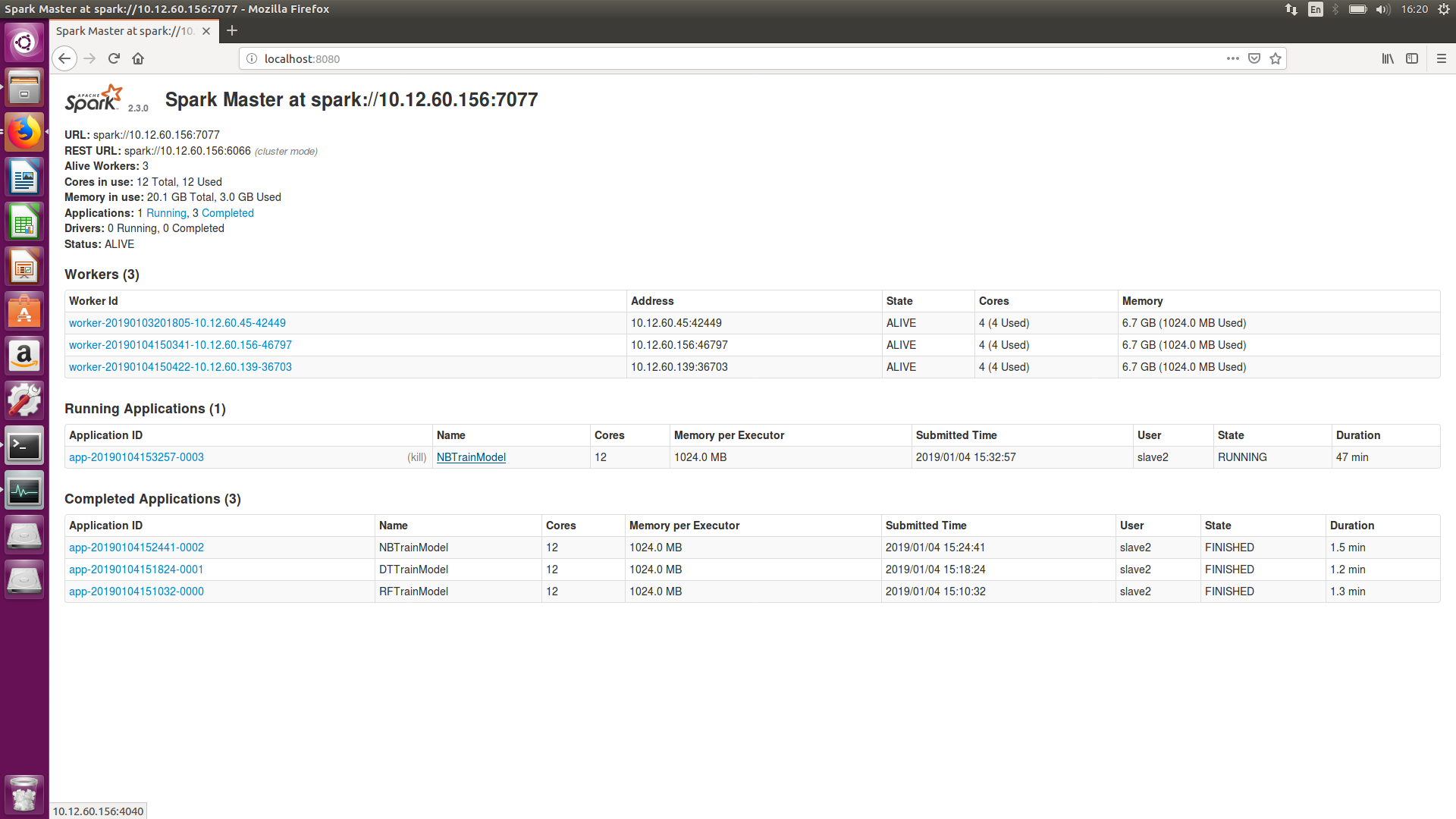}
    \caption{Spark-UI: After submitting a job to testbed}
    \label{fig:sparkJob}
\end{figure*}

\begin{figure*}
    \centering
    \includegraphics[width=0.9\linewidth]{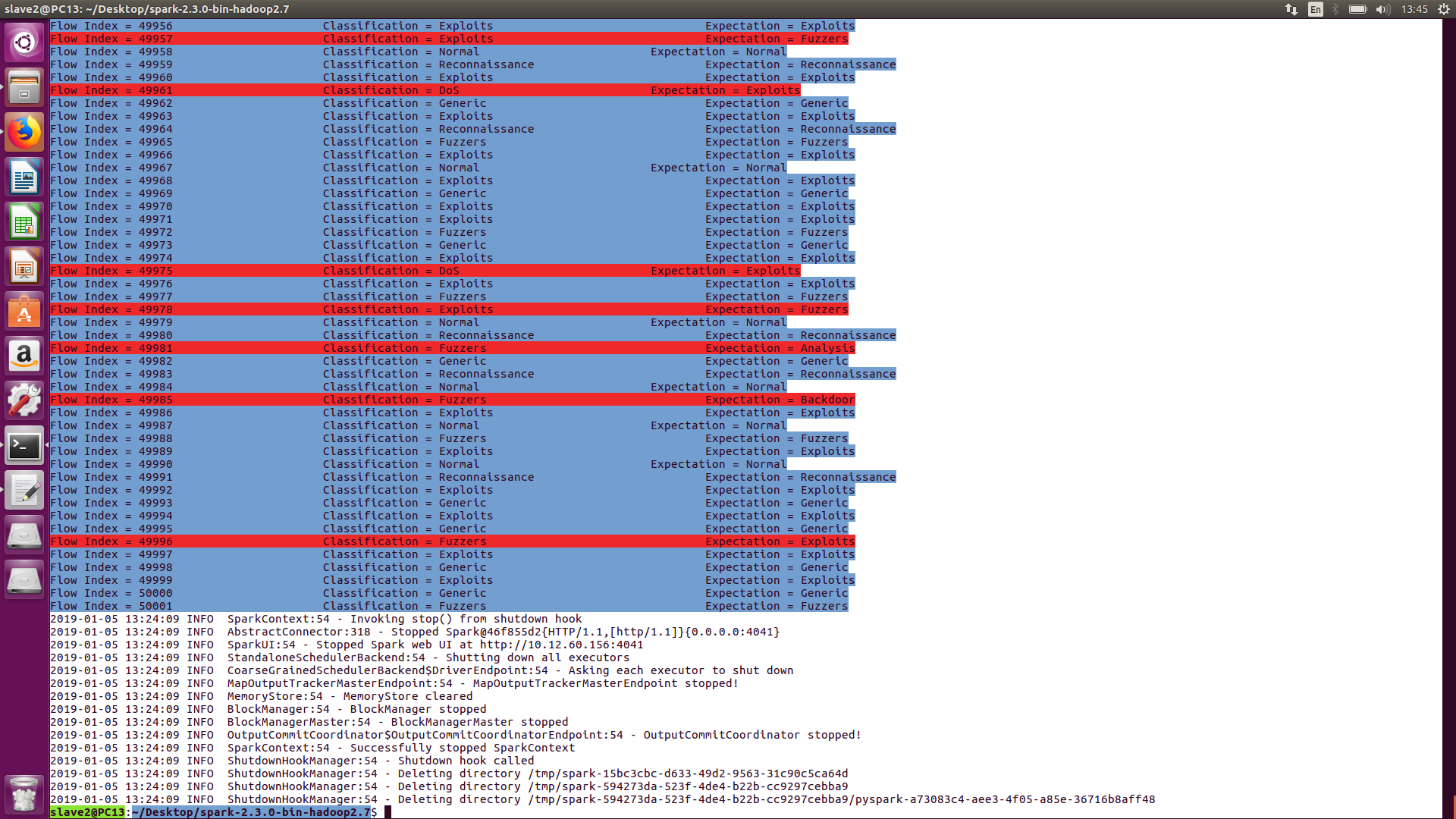}
    \caption{Snapshot view of simulated traffic: SAD-F Result}
    \label{fig:spark-job-end}
\end{figure*}

In addition to the system execution, figure \ref{fig:jobEx}, figure \ref{fig:datPrep}, figure \ref{fig:sparkJob} and figure \ref{fig:spark-job-end} present a step by step result of the test-bed, which includes "view of Spark executor after submitting job to testbed", "preprocessing of data","snapshot of spark user interface after submission of job to testbed" and "view of simulated traffic". 

After submission of spark job one can see it's status including worker node information, spark cluster specification, running jobs and completed job by visiting spark web user interface using REST URL or SPARK URL. 

Spark executor's are the proof that the job is being executing in right manner and it is reflected in figure \ref{fig:jobEx}. Figure-\ref{fig:datPrep} reflects the snapshot of preprocessing work which is quite interesting for detailed working information of the preprocessor module with respect to the model/classifier.

In Figure-\ref{fig:spark-job-end} vulnerable traffic is highlighted with red color to distinguish from normal traffic. At this stage, the proposed framework is not generating any automated notifications for manual action, but this can be extended and incorporated in the framework for further actions.

\section{Conclusion} \label{conclusion}
In this paper, we present SAD-F, a scalable Spark-based live DDoS detection framework that is capable of analyzing potential DDoS attacks with no time delays, as the performance of the framework is tested against live and passive traffic. SAD-F captures live network traffic, preprocesses it to extract relevant information in brief form, and uses ML-Spark algorithms to run detection algorithm for DDoS flooding attacks. SAD-F solves the scalability, memory inefficiency, and process complexity issues of conventional solution by utilizing parallel data processing with low latency and high efficiency.

Before the framework deployment, we tested the dataset on a local machine (8GB of RAM and 2.50 GHz of processor), the evaluation results showed that SAD-F would take maximum time of 12605 sec, i.e. 210 minutes to train and test the overall file for the worst case with 92\% of accuracy with KNN model and it took 6536 sec, i.e 108 minutes for the best case with 72\% of accuracy with NB classifier for the maximum file size of 150MB which comprised of an estimated 1300K packets count. However our suggested SAD-F framework with [low-end] testbed configuration took [5.0001 seconds and 1075.6666 seconds] to process (file loading, preprocessing to classification) for 100K and 3400K packets count respectively with RF classifier as the best case. While, the proposed framework showed best results, according to proof of study, SAD-F framework with [high-end] testbed configuration took [4.3333 seconds and 191.0001 seconds] to process (file loading, preprocessing to detecting) for 100K and 3400K packets count respectively, with DT classifier as the best case.
Based on the framework benchmarks, we conclude active mode of our proposed framework is significantly efficient in terms of preprocessing and DDoS detection than passive mode, which is traffic selector method.

Further, we noticed that data capturing phase consumes more memory than preprocessing phase. Moreover, the appropriate increase in cluster size can increase the lack of CPU utilization, which will improve each stage of this framework. The proposed framework can be used in parallel with conventional security mechanisms to further optimise the results and detection time. 

\newpage
\bibliographystyle{unsrt}
\bibliography{Arxiv,QK} 
\end{document}